\documentclass[epj]{svjour}
\usepackage{graphics}
\usepackage{amsmath}
\usepackage{amssymb}
\newcommand{\vep}{\varepsilon}
\newcommand{\lrangle}[1]{\langle{#1}\rangle}

\newcommand{\rmi}{\textrm{i}}

\newcommand{\refeq}[1]{(\ref{#1})}
\newcommand{\mib}[1]{\ensuremath{\boldsymbol{#1}}}
\begin{document}
\title{
 Electronic Properties Close to Dirac Cone   
  in Two-Dimensional Organic Conductor  $\alpha$-(BEDT-TTF)$_2$I$_3$ 
}
\subtitle{
}
\author{
Shinya Katayama\inst{1}, 
Akito Kobayashi\inst{2}, and 
Yoshikazu Suzumura\inst{1}
}
%
%
\institute{
Department of Physics, Nagoya University, Nagoya 464-8602, Japan \and
Institute for Advanced Research, Nagoya University, Nagoya 464-8602, Japan
}
\date{Received 17 October 2008 / Revised in final form 8 December 2008}
%
\abstract{
A zero-gap state (ZGS) 
  has been found in a bulk system 
 of two-dimensional organic conductor, 
  $\alpha$-(BEDT-TTF)$_2$I$_3$ salt 
   which consists of four sites of donor molecules in a unit cell.  
In the present paper,  
 the characteristic of the ZGS is analyzed in detail and   
   the electronic properties 
   are examined  in the vicinity of the Dirac point where 
     the conduction and valence bands degenerate
     to form the zero-gap. 
The eigenvectors of the energy band 
  have four components of respective sites, 
  where two of them correspond to 
  inequivalent sites and the other two 
  correspond to equivalent sites. 
It is shown that the former exhibits an 
       exotic momentum dependence around the contact point 
      and the latter shows almost a constant dependence. 
The density of states of each site close to the Dirac point 
  is calculated to demonstrate the temperature dependence of 
   the local magnetic susceptibility and 
   the local nuclear magnetic relaxation rate. 
Further, 
   the robust property of the ZGS against 
   the anion potential is also shown by using the second-order perturbation. 
} 
\authorrunning{S. Katayama \textit{et. al.}}
\titlerunning{
 Electronic Properties Close to Dirac Cone   
  in $\alpha$-(BEDT-TTF)$_2$I$_3$ 
}
\maketitle
\section{Introduction}
\label{intro}
Organic conductor, BEDT-TTF \\(bis(ethylene)dithiotetrathiafulvalene) 
  salt has been studied extensively 
  since the various kinds of the transfer energies between 
  BEDT-TTF molecules show 
   exotic phenomena in condensed matter physics \cite{ishiguro}.
Recently, a quasi-two-dimensional conductor, 
 $\alpha$-(BEDT-TTF)$_2$I$_3$ salt \cite{bender,mori1984}, 
  becomes remarkable 
  due to the theoretical finding of a massless Dirac particle under pressures 
  \cite{kobayashi_2004_sc,katayama_zgs}
  based on the experimental data of transfer energies \cite{kondo}. 
Such a particle exhibits  the the zero-gap state (ZGS) 
    where the valence band and the conduction band touch each other 
       at a  momenta, $\pm\mib{k}_0$ in the Brillouin zone, 
       and a Dirac point is produced. 
As the noticeable properties due to transfer energies,  
 the location of $\mib{k}_0$ varies under pressure and 
 the energy band around the Fermi energy is described by  
   a linear dispersion but with  an anisotropic Fermi velocity, i.e., 
    by the tilted Weyl equation \cite{kobayashi_zgs}. 

The intriguing property in the $\alpha$-(BEDT-TTF)$_2$I$_3$ salt 
   appears in the transport phenomena, which  
    come from  a zero-gap conductor.  
The Hall coefficient decreases rapidly with a power law as a function 
of temperature \cite{tajima_2000}, 
  while the electronic conductivity, $\sigma$, 
  stays almost constant ($\sigma\sim h/e^2$ per BEDT-TTF layer) 
  \cite{tajima_2007}. 
Such a behavior of the transport property has been successfully 
  explained in terms of the ZGS \cite{kobayashi_hall,katayama_ele}. 

We note that there are several materials which show the 
  Dirac fermions. 
The Dirac particle in condensed matter physics was 
  first discovered in graphene (single layer graphite) 
  \cite{novoselov,zhang_nature}. 
The graphene shows the isotropic linear dispersion 
  around the corner of the Brillouin zone \cite{wallace,ajiki}. 
The structure of the Landau level \cite{ando,montambaux}, 
  anomalous transport phenomena 
  \cite{andob,zheng,gusynin,fukuyama2007,nomura}, 
  and impurity effect 
  \cite{shonando,kumazaki1,peres,cheianov,wehling}, 
  of the graphene have been proposed. 
Further, the bismuth has a small band gap with anisotropic velocities, 
  and is described by the Dirac equation \cite{wolff}. 
The large diamagnetism appears due to the inter-band 
  effect of magnetic field \cite{fukuyama_bi}. 
The ZGS also occurs in the HgTe quantum well by changing the 
  thickness of the well \cite{zhang}.

There are still several issues on the ZGS of 
  the $\alpha$-(BEDT-TTF)$_2$I$_3$ salt, 
   which are not yet clarified. 
(i) 
It is not obvious  how the ZGS is constructed by  BEDT-TTF molecules 
   with  several transfer energies.  
There are 
  four molecules located at A, A', B, and C sites in a unit cell,  
    where  A and A' are equivalent sites 
           due to the inversion symmetry. 
The features of these three kinds of molecules (A, B, and C) 
  are different with each other. 
There exists the charge disproportionation even at high temperatures 
  \cite{moroto,takano}, and the relation is found as 
     $\lrangle{n_C}>\lrangle{n_A}=\lrangle{n_{A'}}>
  \lrangle{n_B}$ \cite{kakiuchi}, where 
  $\lrangle{n_{\alpha}}$ denotes the amount of the charge. 
The difference in the magnitude of  $\lrangle{n_{\alpha}}$ 
becomes large under pressures \cite{takahashi_pc}. 
Further, it has been found recently that the local 
  magnetic susceptibility  $\chi_{\alpha}$ shows a relation, 
  $\chi_C>\chi_A=\chi_{A'}>\chi_B$ \cite{takahashi_pc}, 
  which has the same relation as that of the charge disproportionation. 
In the present study, we examine the respective contribution of four 
  different molecules to the ZGS, and show that the ZGS 
  does not imply similarity between the susceptibilities and the 
  charge disproportionation. 

(ii) 
It is known that the ZGS is robust against pressure 
 although the location of the contact point, $\pm\mib{k}_0$, 
  varies 
   due to variation of  transfer energies under the  pressure. 
Recently, the ZGS is examined in 
the presence of the anion potential, 
  which gives the local potential on 
    conduction electrons at each molecules \cite{kondo_pre}. 
It is of interest to analyze the effect of the anion potential on 
  the stability of the ZGS   and 
    the location of the contact point. 

In the present paper, 
  we study the role of the respective sites on the ZGS 
  by calculating the local electronic state in the vicinity of the 
  Dirac point. 
In \S \ref{model},  
  the energy bands and charge disproportionation 
  are calculated for the $\alpha$-(BEDT-TTF)$_2$I$_3$ salt 
  by using transfer energies estimated from the first principle calculation.
  \cite{kino_1} 
The perturbation theory is also applied 
  for the analysis of 
  the electronic state close to the Dirac point and anion potential. 
In \S \ref{tilted_weyl_equation}, 
  the electronic state close to the Dirac point 
  is calculated 
  by focusing on the local properties of four donor molecules. 
In \S \ref{anion}, 
  the effect of the anion potential is examined. 
In \S \ref{nmr},  
 the local density of states is calculated and is applied to 
  the estimation of   
 the temperature dependences of magnetic susceptibility and 
   the nuclear magnetic relaxation rate.  
Finally, the summary is given in \S \ref{summary}.

\section{Band calculation}
\label{model}
\begin{figure}
\begin{center}
\resizebox{0.35\textwidth}{!}{
\includegraphics{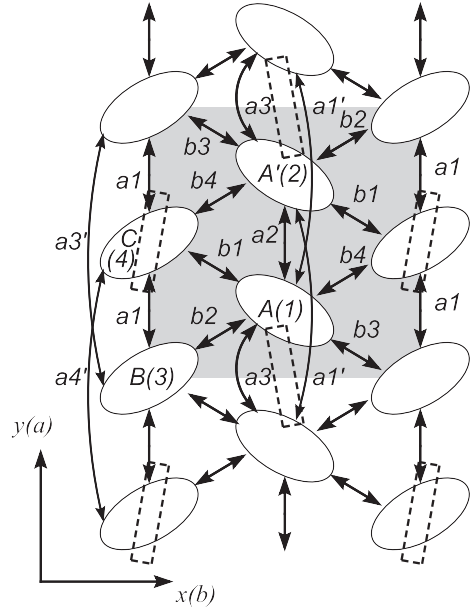}
}
\caption{
Structure of the conducting plane of the $\alpha$-(BEDT-TTF)$_2$I$_3$ salt. 
The ellipse denotes a BEDT-TTF molecule and 
  the unit cell of the salt is drawn by the gray square. 
The bonds, $a1$, $a2$, $a3$, $b1$, $b2$, $b3$, 
  and $b4$, denote the nearest neighbor 
  electron hoppings 
  and $a1'$, $a3'$, and $a4'$ denote the next nearest neighbor ones. 
Anions exist above and below dashed rectangles of
  conducting plane. 
}
\label{unitcell}
\end{center}
\end{figure}
The conducting layer of  $\alpha$-(BEDT-TTF)$_2$I$_3$ is 
    shown in Figure \ref{unitcell}, which are stacked 
    with anion layers alternately. 
The notations of the bonds, 
   $a1$, $a2$, $a3$, $b1$, $b2$, $b3$, and $b4$ in Figure \ref{unitcell}, 
   are the same as those in Ref. \cite{mori1984}, and 
   quantities $a1'$, $a3'$, and $a4'$
   denote those at next nearest neighbor sites, 
   which were introduced by Kino and Miyazaki \cite{kino_1}.  
In the present calculation, 
   we adopt transfer energies evaluated from 
   the first principle calculation 
   at $T=8$ K 
   under the ambient pressure \cite{kino_1}. 
Those in the unit of eV are given as 
  $ t_{a1} = -0.0267$, 
  $ t_{a2} = -0.0511$, 
  $ t_{a3} = 0.0323$, 
  $ t_{b1} = 0.1241$, 
  $ t_{b2} = 0.1296$, 
  $ t_{b3} = 0.0513$, 
  $ t_{b4} = 0.0152$, 
  $ t_{a1'} = 0.0119$, 
  $ t_{a3'} = 0.0046$, and 
  $ t_{a4'} = 0.0060$. 
The unit of the energy is taken as eV, hereafter. 
Other choice of transfer energies is discussed later.

The charge disproportionation in the $\alpha$-type 
   BEDT-TTF salt (Fig. \ref{unitcell}) is examined by
    considering the extended Hubbard model, 
\begin{align}
H&=
\sum^{}_{(i\alpha,j\beta),\sigma}t_{i\alpha:j\beta}
c^{\dag}_{i\alpha\sigma}c_{j\beta\sigma}
+
\sum^{}_{i\alpha,\sigma}I_{\alpha}
c^{\dag}_{i\alpha\sigma}c_{i\alpha\sigma}\nonumber\\
&+
\sum^{}_{i\alpha}U
c^{\dag}_{i\alpha\uparrow}c^{\dag}_{i\alpha\downarrow}
c_{i\alpha\downarrow}c_{i\alpha\uparrow}\nonumber\\
&+
\sum^{}_{n.n.,\sigma\sigma'}V_{i\alpha:j\beta}
c^{\dag}_{i\alpha\sigma}c^{\dag}_{j\beta\sigma'}
c_{j\beta\sigma'}c_{i\alpha\sigma},
\label{hamiltonian}
\end{align}
  where $c_{i\alpha\sigma}$ denotes the annihilation operator of the 
  fermion at the $\alpha$-th site in the $i$-th unit cell with the spin, 
  $\sigma(=\uparrow,\downarrow)$. 
In Figure \ref{unitcell}, 
  sites A, A', B, and C correspond to 
  $1$, $2$, $3$, and $4$, respectively, 
  and $\alpha,\beta=1,\ 2,\ 3$, and $4$.
The quantity 
  $t_{i\alpha:j\beta}$ is the transfer energy between  
  nearest or next nearest sites. 
$U$ is the on-site Coulomb interaction.
$V_{i\alpha:j\beta}$ is the off-site Coulomb interaction between
  $(i,\alpha)$ and $(j,\beta)$ sites, where the interaction is taken as 
  $V_a$ for $a1$, $a2$, and $a3$ bonds, and 
  $V_b$ for $b1$, $b2$, $b3$, and $b4$ bonds. 
The local potentials at A(A'), B, and C sites 
    are given by $I_1(=I_2)$, $I_3$, and $I_4$, respectively,  
    which are  examined in \S \ref{anion}. 
The potential comes from the anion, I$_3$, where the valence is given by 
 $-1$  due to an electron transfer from two BEDT-TTF molecules. 
The anions also form a layer which is located 
  between conducting layers. 
The position projected into the conducting plane (layer) 
  is shown by dashed rectangles in Figure 1 \cite{bender}. 

We treat interactions 
    within the mean-field approximation 
    given by  
\begin{align}
&c^{\dag}_{i\alpha\sigma}c^{\dag}_{j\beta\sigma'}
c_{j\beta\sigma'}c_{i\alpha\sigma}
\nonumber\\
& \rightarrow 
\lrangle{n_{\alpha\sigma}}
c^{\dag}_{j\beta\sigma'}c_{j\beta\sigma'}
+
c^{\dag}_{i\alpha\sigma}c_{i\alpha\sigma}
\lrangle{n_{\beta\sigma'}}-
\lrangle{n_{\alpha\sigma}}\lrangle{n_{\beta\sigma'}}\,,
\end{align}
   where  we take into account only Hartree terms and discard 
   the exchange term. 
The quantity, 
   $\lrangle{c_{i \alpha \sigma}^\dagger c_{i \alpha \sigma}}
   = \lrangle{n_{\alpha \sigma}}$ 
    is calculated self-consistently. 
Using the Fourier transform, 
$c_{i\alpha\sigma}=(1/\sqrt{N})
\sum^{}_{\mib{k}}
c_{\mib{k}\alpha\sigma}
e^{\rmi\mib{k}\cdot\mib{r}_i}$, 
  eq.  \refeq{hamiltonian} is rewritten as 
\begin{gather}
H_{MF}=\sum^{}_{\mib{k},\sigma}
(c_{\mib{k}1\sigma}^{\dag},c_{\mib{k}2\sigma}^{\dag},
c_{\mib{k}3\sigma}^{\dag},c_{\mib{k}4\sigma}^{\dag})\,
(\hat{T}(\mib{k})+\hat{I})\,
\left(
\begin{array}{c}
c_{\mib{k}1\sigma}\\
c_{\mib{k}2\sigma}\\
c_{\mib{k}3\sigma}\\
c_{\mib{k}4\sigma}
\end{array}
\right), 
\label{hmf}
\end{gather}
   where
   $\hat{T}(\mib{k})=[t_{\alpha\beta}(\mib{k})]$, and 
   $\hat{I}=[I_{\alpha}\delta_{\alpha\beta}]$. 
The $4\times4$ matrix elements, $t_{\alpha\beta}(\mib{k})$ 
  ($\alpha,\beta=1,\cdots,4$), are given by  
\begin{align}
t_{11}(\mib{k})&=2t_{a1'}\cos k_y+C_{1}\,,&
t_{12}(\mib{k})&=t_{a2}+t_{a3}e^{-\rmi k_y}\,,\nonumber\\
t_{13}(\mib{k})&=t_{b2}+t_{b3}e^{\rmi k_x}\,,&
t_{14}(\mib{k})&=t_{b1}+t_{b4}e^{\rmi k_x}\,,\nonumber\\
t_{22}(\mib{k})&=2t_{a1'}\cos k_y+C_{2},&
t_{23}(\mib{k})&=t_{b3}e^{\rmi k_y}+t_{b2}e^{\rmi(k_x+k_y)}\,,
\nonumber\\
t_{24}(\mib{k})&=t_{b4}+t_{b1}e^{\rmi k_x}\,,&
t_{33}(\mib{k})&=2t_{a3'}\cos k_y+C_{3}\,,\nonumber\\
t_{34}(\mib{k})&=t_{a1}+t_{a1}e^{-\rmi k_y}\,,&
t_{44}(\mib{k})&=2t_{a4'}\cos k_y+C_{4}\,,\nonumber\\
t_{\alpha\beta}(\mib{k})&=t_{\beta\alpha}(\mib{k})^*\,,&&
\label{matrix4}
\end{align}
  with
\begin{align}
 C_{1}&=
 U\lrangle{n_{1}}/2+2V_a\lrangle{n_{2}}+
2V_b(\lrangle{n_{3}}+\lrangle{n_{4}}), \nonumber\\
 C_{2}&=
U\lrangle{n_{2}}/2+2V_a\lrangle{n_{1}}+
2V_b(\lrangle{n_{3}}+\lrangle{n_{4}}), \nonumber\\
C_{3} &=
U\lrangle{n_{3}}/2+2V_a\lrangle{n_{4}}+
2V_b(\lrangle{n_{1}}+\lrangle{n_{2}}), \nonumber\\
C_{4} &=
U\lrangle{n_{4}}/2+2V_a\lrangle{n_{3}}+
2V_b(\lrangle{n_{1}}+\lrangle{n_{2}}),
\label{diagonal_matrix_element}
\end{align} 
  where 
  $\lrangle{n_{\alpha}}=
  \lrangle{n_{\alpha\uparrow}}+\lrangle{n_{\alpha\downarrow}}$. 
In eq. \refeq{diagonal_matrix_element}, the wave numbers 
  $k_x$ and $k_y$ are measured by $1/a$, where $a$ denotes a lattice 
  constant. 
Note that, in the ZGS, there is no magnetic moment at any site 
  i.e., 
 $\lrangle{m_{\alpha}}
 (=\lrangle{n_{\alpha\uparrow}}-\lrangle{n_{\alpha\downarrow}})=0$. 
The relation, $\lrangle{n_{1}}=\lrangle{n_{2}}$, is satisfied 
  due to the inversion symmetry. 
Quantities $\lrangle{n_{\alpha}}$ and the chemical potential $\mu$ are 
  calculated self-consistently by 
\begin{align}
&\sum^{}_{\mib{k}}(\hat{T}(\mib{k})+\hat{I}-\mu \hat{E})
\mib{d}_{\gamma}(\mib{k})
  =\xi_{\gamma}(\mib{k})\mib{d}_{\gamma}(\mib{k})\,,
  \label{eigen_equation} \\
&\lrangle{n_{\alpha}}=\frac{1}{N}\sum^{}_{\mib{k}\gamma}
                     |d_{\alpha\gamma}(\mib{k})|^2
                        f(\xi_{\gamma}(\mib{k}))\,, \\
&\sum^{}_{\alpha}\lrangle{n_{\alpha}}=6\,, 
\label{nalpha}
\end{align}
  where $f(x)(=1/({\rm e}^{(x-\mu)/T}+1))$ 
  is the Fermi distribution function, and 
  $T$ is a temperature 
  (The unit of $T$ is given as eV, and 
  Boltzmann constant is taken as unity). 
For simplicity, we set $T=0$ and 
  $f(\xi_{\gamma}(\mib{k}))$ in eq. \refeq{nalpha} 
  is replaced by a step function, 
  $\theta(\vep_F-\xi_{\gamma}(\mib{k}))$. 
$\hat{E}$ is the unit matrix. 
The $\alpha$-th component of the vector $\mib{d}_{\gamma}(\mib{k})$ 
  ($\alpha=1,\cdots,4$) is defined 
  as $d_{\alpha\gamma}(\mib{k})$. 
The energy band, $\xi_{\gamma}(\mib{k})$ 
      ($\xi_{1}(\mib{k})\geq\xi_{2}(\mib{k})\geq
       \xi_{3}(\mib{k})\geq\xi_{4}(\mib{k})$), 
      and the eigenvectors 
          $\mib{d}_{\gamma}(\mib{k})$  are 
          obtained  by diagonalizing eq. \refeq{eigen_equation}. 


\begin{figure}
\begin{center}
\resizebox{0.30\textwidth}{!}{
\includegraphics{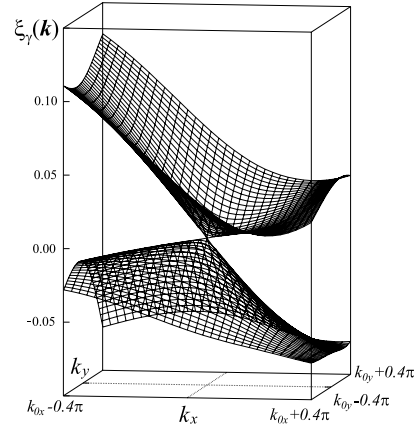}
}
\caption{
Energy band, $\xi_{\gamma}(\mib{k})$ ($\gamma=1,2$), showing 
  Dirac cone in the vicinity of the contact point, 
  $\mib{k}_0=(0.686\pi,-0.443\pi)$.
The Coulomb interactions, $U=0.4$, $V_a=0.17$, $V_b=0.05$, 
  are taken into account in addition to the 
  transfer energies, which are estimated by the first principle 
  calculation at 8 K under ambient pressure. 
We set $\xi_{1}(\mib{k}_0)=\xi_{2}(\mib{k}_0)=0$. 
}
\label{band}
\end{center}
\end{figure}
We choose parameters, $U$, $V_a$, and $V_b$, in order to obtain  
  the observed $\lrangle{n_{\alpha}}$ \cite{kakiuchi},   
  in which 
  the largest electron number is given at the  C site
  and the smallest one is given at the B site.  
In the present calculation, these  parameters are taken  as 
  $U=0.4$, $V_a=0.17$, $V_b=0.05$, and $I_{\alpha}=0.0$ 
  \cite{kobayashi_2005_sc}, which give 
  the ZGS  with  contact points, $\mib{k}_0 = \pm(0.686\pi,-0.443\pi)$. 
Notice that the case $U=V_a=V_b=0$ also gives 
  the ZGS as shown in Ref. \cite{kino_1}. 
For the present case, the energy band of the Dirac cone 
  around the contact point is shown in Figure \ref{band}. 
The charge disproportionation is given by 
  $\lrangle{n_{1}}=\lrangle{n_{2}}=1.460$
   for A and A' sites,  $\lrangle{n_{3}}=1.369$ 
   for B site, and $\lrangle{n_{4}}=1.711$ for C site. 


Next, we examine the electronic state close to $\mib{k}_0$ 
  within the perturbative method. 
Instead of the band index $\gamma$, we use 
   $\zeta=1,2$ for valence and conduction bands 
  and $\eta=3,4$ for the lower two bands. 
The Coulomb interaction and the charge disproportionation are 
   determined unperturbative method,
       i.e., within eqs. \refeq{eigen_equation}-\refeq{nalpha}. 
We construct the effective Hamiltonian as follows:  
\begin{align}
H_{eff}&=H_0+
V+
V\dfrac{1}{E_0-H_0}V+\cdots,
\label{hamiltonian_eff}
\end{align}
  where $V$ is the perturbative Hamiltonian and the explicit 
  form is given in the next section. 
$H_0$ denotes the Hamiltonian, eq. \refeq{hmf}, 
  at $\mib{k}_0$, with the effect of 
  the charge disproportionation 
  ($\lrangle{n_{1}}=\lrangle{n_{2}}=1.460$, 
   $\lrangle{n_{3}}=1.369$, $\lrangle{n_{4}}=1.711$, 
   for $U=0.4$, $V_a=0.17$, $V_b=0.05$, and 
   $\hat{I}=0$). 
Since the system exhibits the ZGS, 
  the upper two eigenvalues of $H_0$ degenerate 
  on the Fermi energy, i.e. 
  $E_0=\xi_{1}(\mib{k}_0)=\xi_{2}(\mib{k}_0)=0$,
  and the lower ones are given as 
  $E_3=\xi_{3}(\mib{k}_0)=-0.218$, and 
  $E_4=\xi_{4}(\mib{k}_0)=-0.404$. 
The eigenstates of $H_0$ described by 
  the Luttinger-Kohn representation \cite{luttinger}
  for $\xi_{\zeta}(\mib{k}_0)$ and 
  $\xi_{\eta}(\mib{k}_0)$ 
  are given by $|\zeta\rangle$ ($\zeta=1,2$) and 
  $|\eta\rangle$ ($\eta=3,4$), respectively. 
Note that $|\zeta\rangle$ can not be determined uniquely 
  due to the degeneracy. 
The detail forms of $\mib{d}_1$ and $\mib{d}_2$ are discussed in the 
  next section. 
On the other hand, $|\eta\rangle$ can be described uniquely. 
Actually, the eigenvectors, $\mib{d}_3(\mib{k}_0)$ and 
   $\mib{d}_4(\mib{k}_0)$ are calculated as 
\begin{align}
\mib{d}_{3}(\mib{k}_0)&=
\left(
\begin{array}{c}
0.5154\\
-0.1944-0.4477\rmi\\
-0.0944+0.4392\rmi\\
-0.2410-0.4571\rmi
\end{array}
\right),\nonumber\\
\mib{d}_{4}(\mib{k}_0)&=
\left(
\begin{array}{c}
0.5190\\
0.3211+0.4077\rmi\\
-0.4332-0.0311\rmi\\
-0.4233+0.3057\rmi
\end{array}
\right). 
\label{eigen_vector34}
\end{align}
When one of the linear combination of $|\zeta\rangle$, 
  $|\psi_{\zeta}\rangle=a_{1\zeta}(\mib{k})|1\rangle+
  a_{2\zeta}(\mib{k})|2\rangle$ is considered, 
  we obtain the equation for the energy 
  ($\Delta E_{\zeta}(\mib{k})$) up to the second order of $V$ given as, 
\begin{align}
&\sum^{}_{\zeta''=1,2}
\left[
\langle\zeta|V|\zeta''\rangle
+
\sum_{\eta=3,4}
\dfrac{\langle\zeta|V|\eta\rangle
\langle\eta|V|\zeta''\rangle}{\xi_{\eta}(\mib{k}_0)}
\right]
a_{\zeta''\zeta'}(\mib{k})\nonumber\\
&=\Delta E_{\zeta}(\mib{k})a_{\zeta\zeta'}(\mib{k}). 
\label{perturbation_equation}
\end{align}
In the next section, we discuss the ZGS using both the 
  exact and perturbative calculations.

\section{Electronic state in the vicinity of the Dirac point}
\label{tilted_weyl_equation}
In \S \ref{tilted_weyl_equation} and \S \ref{anion}, 
  the Dirac point 
  $\mib{k}_0=(0.686\pi,-0.443\pi)$ is taken as 
  the origin of the wave vector ($\mib{k}-\mib{k}_0\rightarrow\mib{k}$).

\begin{figure}
\begin{center}
\resizebox{0.45\textwidth}{!}{
\includegraphics{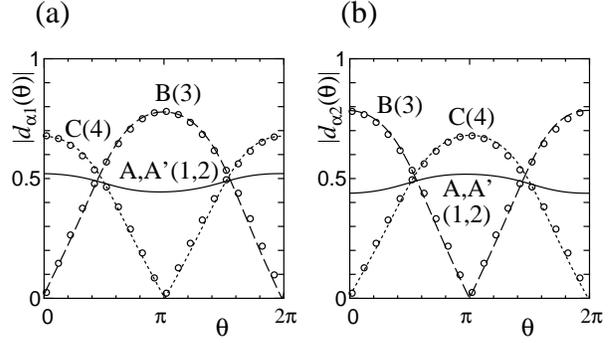}
}
\caption{
$\theta$-dependence of the absolute values of the eigenvectors, 
$|d_{\alpha1}(\theta)|$ (a) and $|d_{\alpha2}(\theta)|$ (b)
($\alpha=A,A',B,C$). 
The open circles are obtained from eq.~(\ref{eigenvector_d}).  
The radius $\delta'(=|\mib{k}-\mib{k}_0|)$ is set as 
$\delta'=10^{-7}\pi$. $\theta$ denotes the angle between 
$\mib{k}-\mib{k}_0$ and $k_x$-axis. 
}
\label{zv_theta}
\end{center}
\end{figure}

First, we discuss the eigenvector with $\mib{k}$ close to $\mib{k}_0$. 
Figure \ref{zv_theta} shows the $\theta$-dependences of 
  $|d_{\alpha1}(\theta)|$ (a) and $|d_{\alpha2}(\theta)|$ (b), which 
  are calculated by diagonalizing eq. \refeq{nalpha}
  ($(k_x,k_y)=\delta'(\cos\theta,\sin\theta)$). 
The notable feature of Figure \ref{zv_theta} (a) is the existence of 
 a   node for the components of both B site and C site, e.g.,  
  $d_{31}(0)=0$ and $d_{41}(\pi)=0$. 
For $\theta\sim0$, $|d_{41}(\theta)|$ is the largest, 
whereas  $|d_{31}(\theta)|$ becomes the largest for $\theta\sim\pi$. 
On the other hand, $|d_{\alpha1}(\theta)|$ for $\alpha=1,2$ 
  has weak $\theta$-dependence. 
It is also seen the relation, 
  $|d_{\alpha1}(\theta)|=|d_{\alpha2}(\theta+\pi)|$.

The behavior of Figure \ref{zv_theta} is analyzed by the 
  tilted Weyl equation. 
For small $|\mib{k}|$, the Hamiltonian, $V$, is written as 
\begin{align}
V(\mib{k})&=
|\mib{k}_0,\alpha\rangle
\left[
\mib{k}\cdot
\dfrac{\partial \hat{T}}{\partial \mib{k}}(\mib{k}_0)
\right]\langle\mib{k}_0,\beta|\,,
\label{v1}
\end{align}
  where $|\mib{k},\alpha\rangle$ is the electronic state at site 
  $\alpha(=1,\cdots,4)$ with the wave number $\mib{k}$.  
By substituting eq. \refeq{v1} into eq. \refeq{perturbation_equation} 
  and discarding the second order term, 
  the tilted Weyl equation is written as 
\begin{align}
&
\mib{k}\cdot
\left(
\begin{array}{cc}
  \tilde{\mib{v}}_{11}&\tilde{\mib{v}}_{12}\\
  \tilde{\mib{v}}_{21}&\tilde{\mib{v}}_{22}
\end{array}
\right)
\left(
\begin{array}{c}
a_{1\zeta}(\mib{k})\\a_{2\zeta}(\mib{k})
\end{array}
\right)\nonumber\\
=&
\mib{k}\cdot
\left(
\begin{array}{cc}
  \mib{v}_0+\mib{v}_3&\mib{v}_1-\rmi\mib{v}_2\\
  \mib{v}_1+\rmi\mib{v}_2&\mib{v}_0-\mib{v}_3
\end{array}
\right)
\left(
\begin{array}{c}
a_{1\zeta}(\mib{k})\\a_{2\zeta}(\mib{k})
\end{array}
\right)\nonumber\\
=&\sum^{3}_{\rho=0}
\mib{k}\cdot\mib{v}_{\rho}\sigma_{\rho}
\left(
\begin{array}{c}
a_{1\zeta}(\mib{k})\\a_{2\zeta}(\mib{k})
\end{array}
\right)
=
\Delta E_{\zeta}
\left(
\begin{array}{c}
a_{1\zeta}(\mib{k})\\a_{2\zeta}(\mib{k})
\end{array}
\right). 
\label{tilted}
\end{align}
The velocity, $\tilde{\mib{v}}_{\gamma\gamma'}$ 
  ($\gamma,\gamma'=1,\cdots, 4$) is given by 
\begin{align}
\tilde{\mib{v}}_{\gamma\gamma'}&=
\sum^{}_{\alpha\beta}
\tilde{d}^{*}_{\alpha\gamma}\,
\frac{\partial t_{\alpha\beta}}{\partial\mib{k}}(\mib{k}_0)\,
\tilde{d}_{\beta\gamma'}, 
\label{def_velocity2}
\end{align}
   where $\tilde{d}_{\alpha\gamma}$ is defined as 
   $\tilde{d}_{\alpha\zeta}=d_{\alpha\zeta}(\mib{k}'_0)$ ($\zeta=1,2$) 
   and 
   $\tilde{d}_{\alpha\eta}=d_{\alpha\eta}(\mib{k}_0)$ ($\eta=3,4$). 
Since the components of the eigenvectors of 
  the conduction ($d_{\alpha1}$) 
  and valence ($d_{\alpha2}$) bands 
  can not be determined at $\mib{k}_0$, 
  the vector 
  $\mib{k}'_0=(\delta''\cos\theta,\delta''\sin\theta)$ is introduced,
  where $\delta''$ is an infinitesimally small quantity and $\theta$ is 
  the angle between $\mib{k}'_0$ and $k_x$-axis. 
Velocities, $\mib{v}_{0}$, $\mib{v}_{1}$, $\mib{v}_{2}$, and 
  $\mib{v}_{3}$, are expressed as
  $\mib{v}_{0}=(\tilde{\mib{v}}_{11}+\tilde{\mib{v}}_{22})/2$, 
  $\mib{v}_{1}=\mathrm{Re}(\tilde{\mib{v}}_{12})$, 
  $\mib{v}_{2}=-\mathrm{Im}(\tilde{\mib{v}}_{12})$, and 
  $\mib{v}_{3}=(\tilde{\mib{v}}_{11}-\tilde{\mib{v}}_{22})/2$, and  
  $\sigma_0$ and $\sigma_i$ ($i=1, 2 ,3$) denote unit matrix and 
  Pauli matrix, respectively. 
The eigenvalue of eq.~\refeq{tilted} is calculated as 
\begin{align}
&\Delta E_{1,2}(\mib{k})=
\mib{k}\cdot\mib{v}_0\pm
\sqrt{\sum^{3}_{\rho=1}(\mib{k}\cdot\mib{v}_{\rho})^2}
\nonumber\\
=&
v_{0x}k_x+v_{0y}k_y\pm
\sqrt{V_x^2 k_x^2+V_y^2 k_y^2+
2V_{xy} k_xk_y},
\label{linear_dispersion}
\end{align}
  with 
  $V_x^2 = v_{1x}^2+v_{2x}^2+v_{3x}^2$, 
  $V_y^2 = v_{1y}^2+v_{2y}^2+v_{3y}^2$, and 
  $V_{xy} = v_{1x}v_{1y}+v_{2x}v_{2y}+v_{3x}v_{3y}$. 
It should be noted that 
   $\mib{v}_1$ and $\mib{v}_2$ 
         depend on the wave number $\mib{k}{'_0}$, 
         whereas 
   the velocity vectors  $\mib{v}_0$ and $\mib{v}_3$ and 
  quantities, $V_x^2$, $V_y^2$, and $V_{xy}$ are 
   independent of the choice of the base. 

\begin{figure}
\begin{center}
\resizebox{0.35\textwidth}{!}{
\includegraphics{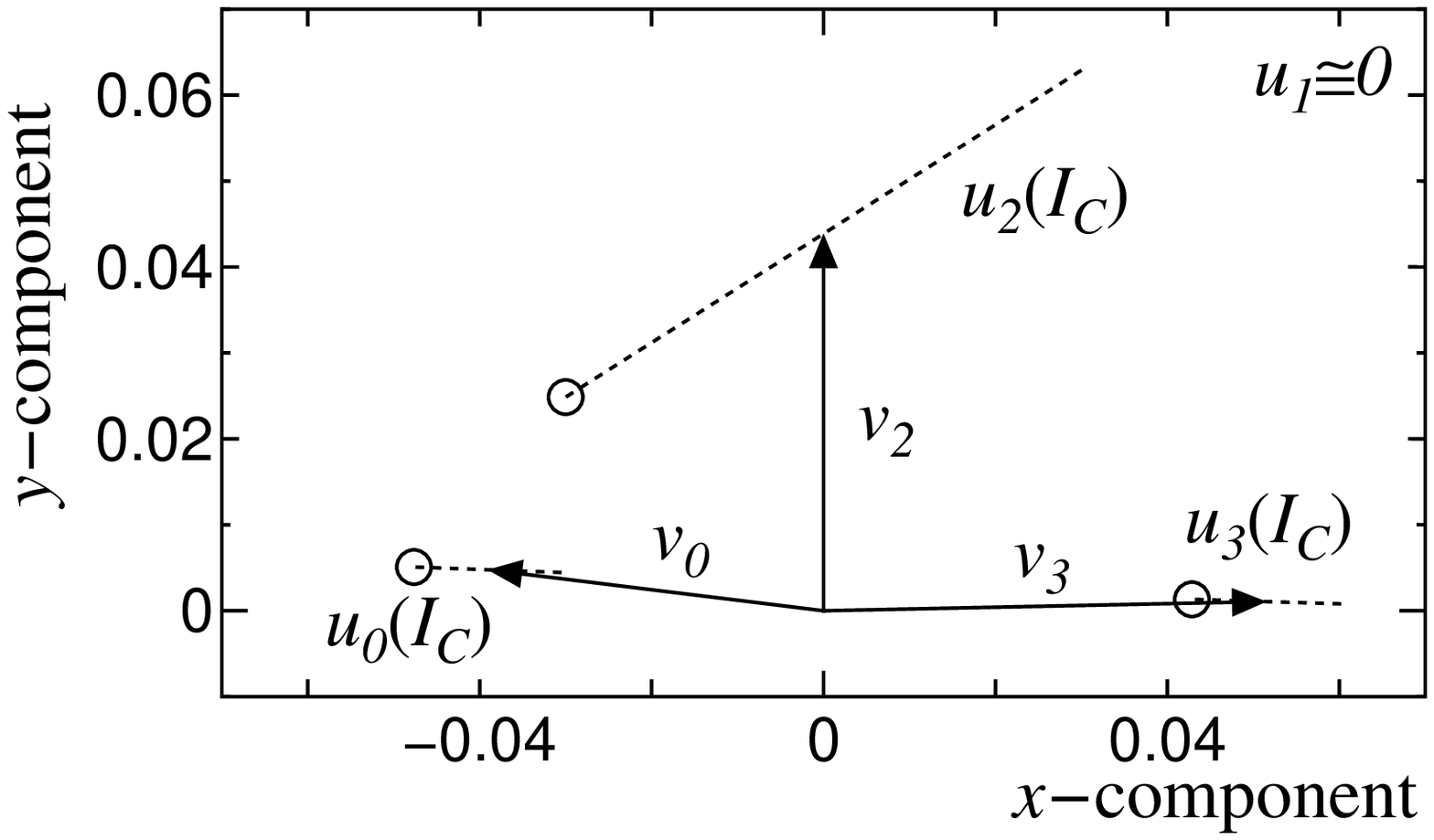}
}
\caption{
Velocity vectors, $\mib{v}_{\rho}$ (eq. \refeq{velocity_const}) 
  and $\mib{u}_{\rho}$ (eq. \refeq{velocity_idep}), 
  where the bases are given by eq. \refeq{eq:11}.
$\mib{v}_{\rho}$ are shown by arrows. 
For $I_A=I_B=0$, and $-0.2<I_C<0.2$, $\mib{u}_{\rho}$ ($\rho=0,2,3$) 
  are shown by dotted line where open circles 
  correspond to $I_C=0.2$. 
$\mib{u}_1$ is small compared with other velocities 
  ($|\mib{u}_1|\sim10^{-3}$). 
}
\label{velocity_fig}
\end{center}
\end{figure}

We choose  $d_{\alpha\gamma}$ as  
 the eigenvector of  $t_{\alpha\beta}(\mib{k}_0)$ with $\theta=0$,
  i.e.,  
\begin{align}
\mib{x}_{1}&=\mib{d}_{1}=
\left(
\begin{array}{c}
0.5194\\
-0.4429+0.2713\rmi\\
0\\
0.6623+0.1473\rmi
\end{array}
\right)\,,
\nonumber\\ 
\mib{x}_{2}&=\mib{d}_{2}=
\left(
\begin{array}{c}
0.4419\\
0.3703-0.2411\rmi\\
0.6125-0.4841\rmi\\
0
\end{array}
\right)\,.
\label{eq:11}
\end{align}
The velocity vectors are given as 
\begin{align}
 \mib{v}_0&=(-0.0389, 0.0048),\ \ 
 \mib{v}_1=(0, -0.0005),\nonumber\\ 
 \mib{v}_2&=(0, 0.0439),\ \ 
 \mib{v}_3=(0.0515, 0.0009). 
\label{velocity_const}
\end{align}
These vectors, 
  are drawn in Figure \ref{velocity_fig}. 
In this case, the effective Hamiltonian can be approximately 
  written as
\begin{align}
H'&=
\left(
\begin{array}{cc}
v_{0x}k_x+v_{0y}k_y+vk_x&-\rmi v'k_y\\
\rmi v'k_y&v_{0x}k_x+v_{0y}k_y-vk_x
\end{array}
\right)\,, 
\label{h_eff2}
\end{align}
  where $v_{0x}=-0.0389$, $v_{0y}=0.0048$, 
  $v'=v_{2y}=0.0439$, and $v=v_{3x}=0.0515$. 
Note that the effective Hamiltonian of 
  eq. \refeq{h_eff2} differs from that of 
  Ref. \cite{kobayashi_hall}, 
  since the choice of $\mib{k}_0'$, i.e. the bases at $\mib{k}_0$, 
  are not the same. 
However, the result does not depend on such a choice.  
The eigenvalue ($\Delta E_{\zeta}$) and eigenvector 
  ($a_{\zeta\zeta'}$) of eq. \refeq{h_eff2} are given as 
\begin{align}
&\Delta E_{1}=\omega=v_{0x}k_x+v_{0y}k_y+\sqrt{v^2k_x^2+v{'}^{2}k_y^2},
\nonumber\\
&\Delta E_{2}=\omega'=v_{0x}k_x+v_{0y}k_y-\sqrt{v^2k_x^2+v{'}^{2}k_y^2},
\label{ev_simple}
\\
&\left(
\begin{array}{c}
a_{11}(\theta)\\a_{21}(\theta)
\end{array}
\right)
=
\left(
\begin{array}{c}
\cos(\theta/2)\\
\rmi\sin(\theta/2)
\end{array}
\right),\nonumber\\
&\left(
\begin{array}{c}
a_{12}(\theta)\\a_{22}(\theta)
\end{array}
\right)
=
\left(
\begin{array}{c}
\sin(\theta/2)\\
-\rmi\cos(\theta/2)
\end{array}
\right),
\label{matrix2ev}
\end{align}
  where the relation $v\simeq v'$ is used for 
    the calculation of $a_{\zeta\zeta'}(\mib{k})$, for simplicity. 
From eqs.~\refeq{eq:11} and \refeq{matrix2ev}, 
the $\theta$-dependence of $d_{\alpha\gamma}$ ($\gamma=1,2$) 
is obtained as  
\begin{align}
\left(
\begin{array}{c}
\mib{d}_{1}(\theta)\\
\mib{d}_{2}(\theta)
\end{array}
\right)
&=
\left(
\begin{array}{cc}
a_{11}(\theta)&a_{12}(\theta)\\
a_{21}(\theta)&a_{22}(\theta)
\end{array}
\right)
\left(
\begin{array}{c}
\mib{x}_{1}\\
\mib{x}_{2}
\end{array}
\right). 
\label{eigenvector_d}
\end{align}
This gives 
 $|d_{41}(\theta)| = |d_{32}(\theta)| \simeq 0.678  |\cos (\theta/2)|$ and 
 $|d_{31}(\theta)| = |d_{42}(\theta)| \simeq 0.785 |\sin (\theta/2)|$ 
 which reproduce well the exact one as shown in Figure~\ref{zv_theta}
 (open circle) within the accuracy of $\sim0.03$. 

\begin{figure}
\begin{center}
\resizebox{0.35\textwidth}{!}{
\includegraphics{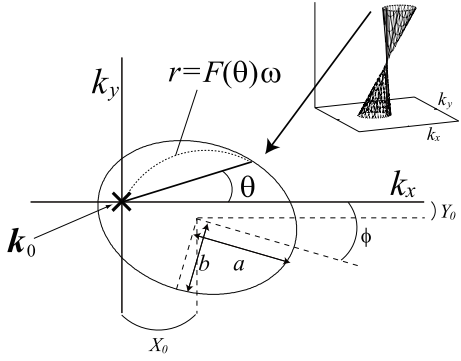}
}
\end{center}
\caption{
Ellipsoidal curve obtained from the condition, 
$\xi=\omega$, for the Dirac cone. $\theta$ is the angle 
between $\mib{k}-\mib{k}_0$ and $k_x$-axis, 
where $\mib{k}_0$ is the point 
on the ellipsoidal curve. 
The inset denotes a tilted Dirac cone where $\omega=0$
 corresponds to the apex, i.e., the Dirac point. 
}
\label{ellipse_fig}
\end{figure}

Next, we discuss the detail structure of the Dirac cone. 
Equation \refeq{ev_simple}  is rewritten as 
\begin{align}
c_1k_x^2+c_2k_y^2+c_3k_xk_y+c_4k_x+c_5k_y&=1 ,
\label{c5}
\end{align}
 where the coefficients are given by  
 $ c_1  = (v^2- v_{0x}^2)/\omega^2$, 
 $ c_2  = (v{'}^2- v_{0y}^2)/\omega^2$, 
 $ c_3  = 2v_{0x}v_{0y}/\omega^2$, 
 $ c_4  = -2 v_{0x}/\omega$ and  
 $ c_5  =2 v_{0y}/\omega$. 
Equation (\ref{c5}) represents the general form  
     of an ellipse as shown  in  Figure \ref{ellipse_fig}. 
The angle, $\phi$, between the long axis of the ellipse and $k_x$-axis, 
  and the ratio, $a/b$, 
  between the long and short axes are calculated as 
\begin{align}
\phi&=\frac{1}{2}\tan^{-1}
\frac{c_3}{c_1-c_2},\nonumber\\
\frac{a}{b}& 
=\sqrt{
 \frac
 { c_1+c_2 + \sqrt{(c_1-c_2)^2 +c_3^2}}
  { c_1+c_2 - \sqrt{(c_1-c_2)^2 +c_3^2}
 }},
\end{align}
  respectively. 
In the present case, one obtains 
     $\phi < 0$ and $a > b$.  
Further, by introducing a polar coordinate, $k_x=r\cos\theta$ and 
  $k_y=r\sin\theta$,   eq. \refeq{c5} is rewritten as 
\begin{align}
r   = \frac{\sqrt{4A_1 + A_2^2} -A_2}{2A_1} 
 \equiv  F(\theta)\omega , 
 \label{eq:16}
\end{align}
where 
 $  A_1 =c_1\cos^2\theta+c_2\sin^2\theta+c_3\cos\theta\sin\theta$ and 
 $A_2 = c_4\cos\theta+c_5\sin\theta$. 
 For  $U=0.4$, $V_a=0.17$, $V_b=0.05$,  
  the $\theta$-dependence of $F(\theta)$ is drawn in Figure \ref{ellipse_fig}, 
  where parameters are given as 
  $a/\omega=48.00$, $b/\omega=35.20$, $X_0/\omega=36.09$, $Y_0/\omega=-7.082$, 
   and   $\phi= - 0.089\pi$ 
   (($X_0$,$Y_0$) denotes the center coordinate of the ellipse.). 
  The general form in terms of  $X_0$ and $Y_0$ is shown  in Appendix.

Let us compare the present result with that of the graphene. 
For the latter case, the Hamiltonian of eq. (\ref{tilted}) 
  is rewritten as  
  $H'=v (k_x\sigma_1+k_y\sigma_2)$,
  which denotes the isotropic Dirac cone. 
The absolute value of the component of the eigenvector is 
  independent of  $\theta$, i.e. 
  $|d_{\alpha\gamma}(\theta)|=1/\sqrt{2}$. 
The eigenvector of the latter case 
  may correspond to the component of A and A' of the former case, 
  in which the $\theta$-dependence is small as shown 
  in Figure \ref{zv_theta}. 
Thus it is found that the existence of the  B and C components 
  of the present case is 
  the characteristic of the Dirac particle of 
  organic conductor, $\alpha$-(BEDT-TTF)$_2$I$_3$ salt, in which 
  there are four sites in the unit cell. 
Further the vanishing of the B and C components at a certain value of 
   $\theta$ seems to be general and exist for other case of 
   transfer energies as discussed in \S \ref{summary}.

\section{The effect of the anion potential}
\label{anion}
Next, we examine the anion potentials, $I_{A}(=I_{1}=I_{2})$, 
 $I_{B}(=I_{3})$, and $I_{C}(=I_{4})$ perturbatively. 
Although the charge disproportionation depends on 
  the anion potential \cite{katayama_proc}, 
  quantities of $\lrangle{n_{\alpha}}$ are fixed to 
  $\lrangle{n_{1}}=\lrangle{n_{2}}=1.460$, 
   $\lrangle{n_{3}}=1.369$, and $\lrangle{n_{4}}=1.711$, which is 
   obtained by the condition, $\hat{I}=0$. 
Such a treatment can be justified for the qualitative discussion 
  of the location of $\mib{k}_0$ since the symmetry relation, 
  $\lrangle{n_{1}}=\lrangle{n_{2}}$, remains due to $I_1=I_2$.

When the perturbative Hamiltonian is given as 
\begin{align}
V(\mib{k})&=
|\mib{k}_0,\alpha\rangle
\left[\mib{k}\cdot
\dfrac{\partial \hat{T}}{\partial \mib{k}}(\mib{k}_0)+\hat{I}
\right]
\langle\mib{k}_0,\beta|,
\label{v2}
\end{align}
  eq. \refeq{perturbation_equation} is rewritten as 
\begin{align}
\sum^{}_{\zeta''=1,2}
[
-\tilde{K}_{\zeta\zeta''}
+\mib{k}\cdot\tilde{\mib{u}}_{\zeta\zeta''}
+\tilde{J}_{\zeta\zeta''}]
a_{\zeta''\zeta'}(\mib{k})
=\Delta E_{\zeta}(\mib{k})a_{\zeta\zeta'}(\mib{k})\,,  
\label{eigen_equation_i}
\end{align}
  with
\begin{gather}
\tilde{K}_{\zeta\zeta'}=
\sum^{}_{\eta=3,4}
\dfrac{(\mib{k}\cdot\tilde{\mib{v}}_{\zeta\eta})
(\mib{k}\cdot\tilde{\mib{v}}_{\eta\zeta'})}
{\xi_{\eta}(\mib{k}_0)}\,,
\nonumber\\
\tilde{\mib{u}}_{\zeta\zeta'}=
\tilde{\mib{v}}_{\zeta\zeta'}-
\sum^{}_{\eta=3,4}
\dfrac{\tilde{\mib{v}}_{\zeta\eta}
\tilde{I}_{\eta\zeta'}+
\tilde{I}_{\zeta\eta}
\tilde{\mib{v}}_{\eta\zeta'}}
{\xi_{\eta}(\mib{k}_0)}\,,\nonumber\\
\tilde{J}_{\zeta\zeta'}=
\tilde{I}_{\zeta\zeta'}-
\sum^{}_{\eta=3,4}
\dfrac{\tilde{I}_{\zeta\eta}
\tilde{I}_{\eta\zeta'}}{\xi_{\eta}(\mib{k}_0)}, 
\label{hamiltonian_second_order}
\end{gather}
  where the matrix elements, 
  $\tilde{I}_{\gamma\gamma'}$ $(\gamma=\zeta,\eta=1,\cdots,4)$ 
  is defined by 
\begin{align}
\tilde{I}_{\gamma\gamma'}&=
\sum^{}_{\alpha}\tilde{d}^{*}_{\alpha\gamma}I_{\alpha}
\tilde{d}_{\alpha\gamma'}. 
\label{hamiltonian_second_order}
\end{align}

\begin{figure}
\begin{center}
\resizebox{0.35\textwidth}{!}{
\includegraphics{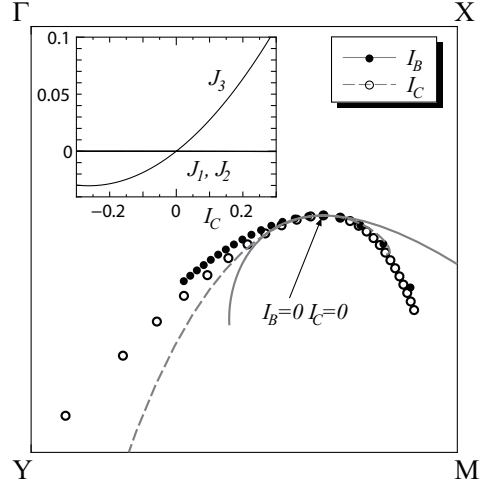}
}
\caption{
Trajectory of the contact point in the quarter part 
  of the Brillouin zone ($0<k_x<\pi$, $-\pi<k_y<0$). 
Filled (open) circle denotes the position of $\mib{k}_0$ 
  calculated from eq. \refeq{eigen_equation} by the condition 
  $I_A=0,-0.3<I_B<0.3,I_C=0$ ($I_A=0,I_B=0,-0.3<I_C<0.3$). 
The corresponding result obtained from eq. \refeq{k0_perturbation} 
  is drawn by solid (dashed) line. 
The inset shows the quantities $J_{1}$, $J_{2}$, and $J_{3}$ 
  as functions of $I_C$. 
}
\label{contact_point_perturbation}
\end{center}
\end{figure}

Discarding the first term of eq. \refeq{eigen_equation_i}
  for simplicity, 
  i.e. $\tilde{K}_{\gamma\gamma'}\rightarrow0$, 
  the energy bands are calculated as 
\begin{align}
\Delta E_{1,2}(\mib{k})&=
\mib{k}\cdot\mib{u}_0+J_0
\pm\sqrt{\sum^{3}_{\rho=1}(\mib{k}\cdot\mib{u}_{\rho}+J_{\rho})^2},
\end{align}
  where $\mib{u}_0=(\tilde{\mib{u}}_{11}+\tilde{\mib{u}}_{22})/2$, 
  $\mib{u}_1=\mathrm{Re}(\tilde{\mib{u}}_{12})$, 
  $\mib{u}_2=-\mathrm{Im}(\tilde{\mib{u}}_{12})$, 
  $\mib{u}_3=(\tilde{\mib{u}}_{11}-\tilde{\mib{u}}_{22})/2$, 
  $J_0=(\tilde{J}_{11}+\tilde{J}_{22})/2$, 
  $J_1=\mathrm{Re}(\tilde{J}_{12})$, 
  $J_2=-\mathrm{Im}(\tilde{J}_{12})$, and 
  $J_3=(\tilde{J}_{11}-\tilde{J}_{22})/2$. 
Compared with the previous section, 
 the velocity is changed by the anion potential. 
Actually, for the bases chosen as 
  eqs. \refeq{eigen_vector34} and \refeq{eq:11}, 
  components of $\mib{u}_{\rho}$ ($\rho=0,\cdots, 3$) are given by 
\begin{align}
u_{\rho x}&=v_{\rho x}+K^A_{\rho x}I_A+
K^B_{\rho x}I_B+K^C_{\rho x}I_C,\nonumber\\
u_{\rho y}&=v_{\rho y}+K^A_{\rho y}I_A+
K^B_{\rho y}I_B+K^C_{\rho y}I_C
\label{velocity_idep}
\end{align}
  where coefficients of $K^{\alpha}_{\rho x}$, $K^{\alpha}_{\rho y}$ 
  are given in Table \ref{tab:1}. 
The velocity vector 
  as a function of $I_C$  is drawn in 
  Figure \ref{velocity_fig}, 
  where  $-0.2<I_C<0.2$ and 
   the other anion potentials are set as $I_A=I_B=0$. 
In the presence of $J_{\rho}$,  the position of the 
  contact point changes from that of the state with $\hat{I}=0$. 
The quantities $J_{\rho}$ ($\rho=0,1,2,3$), 
  which are expanded up to a quadratic term 
  of the anion potential, are described by 
\begin{align}
J_{\rho}=&L^{A}_{\rho}I_A+L^{B}_{\rho}I_B+L^{C}_{\rho}I_C
+L^{AA}_{\rho}I_A^2+L^{BB}_{\rho}I_B^2+L^{CC}_{\rho}I_C^2\nonumber\\
&+L^{AB}_{\rho}I_AI_B+L^{CC}_{\rho}I_BI_C+L^{CA}_{\rho}I_CI_A\,,
\end{align}
  where $L^{\alpha}_{\rho}$ is given in Table \ref{tab:2}. 
The $I_C$-dependences of $J_{1}$, $J_{2}$, $J_{3}$
  ($I_A=I_B=0$) are shown in the 
  inset of Figure \ref{contact_point_perturbation}. 
Note that $J_1$ and $J_2$ are negligibly small compared with $J_3$ 
  (and also for case of the $I_B$-dependence 
  with $I_A=I_C=0$). 
It should be noted that the effect of $J_0$ being the same order of $J_3$
 does not contribute to  the shape of the Dirac cone. 

\begin{table}
\caption{Quantities of coefficients, $K^{\alpha}_{\rho x}$, 
$K^{\alpha}_{\rho y}$ ($\alpha=A,B,C$, $\rho=0,1,2,3$).}
\label{tab:1} 
\begin{center}
\begin{tabular}{ccccc}
\hline\noalign{\smallskip}
$K^{\alpha}_{\rho x}$ & $\rho=0$ & $\rho=1$ & $\rho=2$ & $\rho=3$ \\
\noalign{\smallskip}\hline\noalign{\smallskip}
A & -0.03 & 0.00 &0.01&0.12\\
B & 0.08 & 0.00 &0.14&-0.07\\
C & -0.04 & 0.00 &-0.15&-0.04\\
\noalign{\smallskip}\hline\hline\noalign{\smallskip}
$K^{\alpha}_{\rho y}$ & $\rho=0$ & $\rho=1$ & $\rho=2$ & $\rho=3$ \\
\noalign{\smallskip}\hline\noalign{\smallskip}
A & 0.01 & 0.00 &0.07&0.00\\
B & 0.01 & 0.00 &0.02&0.00\\
C & 0.00 & 0.00 &-0.10&0.00\\
\noalign{\smallskip}\hline
\end{tabular}
\end{center}
\end{table}
\begin{table}
\caption{Quantities of coefficients, $L^{\alpha}_{\rho}$ 
 ($\alpha=A, B, C, AA, BB, CC, AB, BC, CA$).}
\label{tab:2} 
\begin{center}
\begin{tabular}{ccccc}
\hline\noalign{\smallskip}
& $\rho=0$ & $\rho=1$ & $\rho=2$ & $\rho=3$ \\
\noalign{\smallskip}\hline\noalign{\smallskip}
$L^{A}_{\rho}$ & 0.47 & 0.00 &-0.01&0.07\\
$L^{B}_{\rho}$ & 0.30 & 0.00 &0.01&-0.30\\
$L^{C}_{\rho}$ & 0.23 & 0.00 &0.00&0.23\\
\noalign{\smallskip}\hline\noalign{\smallskip}
$L^{AA}_{\rho}$ & 0.86 & 0.00 &-0.26&0.01\\
$L^{BB}_{\rho}$ & 0.42 & 0.00 &0.01&-0.42\\
$L^{CC}_{\rho}$ & 0.44 & 0.00 &0.00&0.44\\
\noalign{\smallskip}\hline\noalign{\smallskip}
$L^{AB}_{\rho}$ & -0.85 & 0.00 &0.25&0.85\\
$L^{BC}_{\rho}$ & 0.00 & 0.00 &-0.27&0.00\\
$L^{CA}_{\rho}$ & -0.87 & 0.00 &0.27&-0.87\\
\noalign{\smallskip}\hline
\end{tabular}
\end{center}
\end{table}

Using the above results, 
  the energy difference between $\Delta E_1(\mib{k})$ and 
  $\Delta E_2(\mib{k})$ is approximately given by 
\begin{align}
&\Delta E_1(\mib{k})-\Delta E_2(\mib{k})\nonumber\\
&\simeq
2\sqrt{(u_{2x}k_x+u_{2y}k_y)^2+
(u_{3x}k_x+u_{3y}k_y+J_3)^2}. 
\end{align}
Then, the contact point, which is given by 
   $\Delta E_1-\Delta E_2=0$, 
   is obtained at 
\begin{align}
k_{0x}=\dfrac{J_3u_{2y}}{u_{2x}u_{3y}-u_{3x}u_{2y}},\ \ 
k_{0y}=\dfrac{-J_3u_{2x}}{u_{2x}u_{3y}-u_{3x}u_{2y}}. 
\label{k0_perturbation}
\end{align}
The trajectories of the contact point calculated from both 
  eqs. \refeq{eigen_equation} (circle) and 
  \refeq{k0_perturbation} (line) 
  are drawn in the main part of 
  Figure \ref{contact_point_perturbation}. 
In case of the exact calculation, it moves from Y point to M point 
  as shown in Ref. \cite{katayama_proc}. 
The trajectory obtained from eq. \refeq{k0_perturbation}
    agrees   qualitatively   with that  of 
       eq. \refeq{eigen_equation} for the small potential.
Thus, one finds the ZGS is robust against the anion potential.


\section{Local density of states and magnetic properties}
\label{nmr}
\begin{figure}
\begin{center}
\resizebox{0.4\textwidth}{!}{
\includegraphics{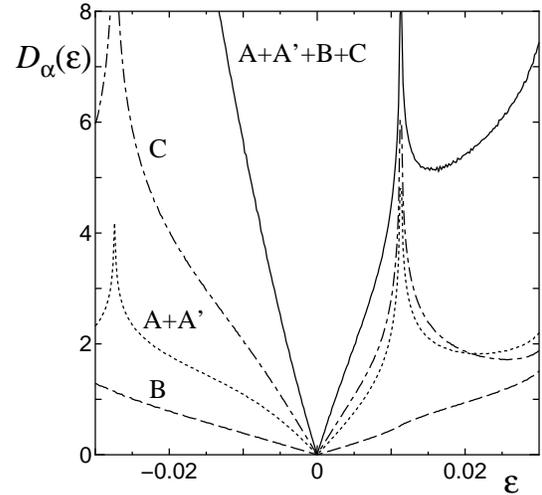}
}
\caption{
Local density of states $D_{\alpha}(\varepsilon)$ per unit cell 
 for A and A' sites (dotted line), B site 
(dashed line), and C site (dot-dashed line), which are 
calculated from eq. \refeq{dos_site}. 
Solid line is the total density of states. 
The parameters of interactions  are taken as 
  $U =0.4$, $V_a=0.17$ and $V_b=0.05$. 
}
\label{dos_site_fig}
\end{center}
\end{figure}

By using the energy band and eigenvector 
  (eq. \refeq{eigen_equation}), 
  the density of states of the $\alpha$ site in a unit cell 
  is calculated as 
\begin{align}
D_{\alpha}(\vep)&=\dfrac{1}{N}\sum^{}_{\mib{k}\gamma}
|d_{\alpha\gamma}(\mib{k})|^2\delta(\vep-\xi_{\gamma}(\mib{k})), 
\label{dos_site}
\end{align}
   where $\alpha=1$(A), $2$(A'), $3$(B), and $4$(C) 
   denote the respective sites. 
The total value of the density of states is given by  
   $\int \mathrm{d}\vep\sum^{4}_{\alpha=1}D_{\alpha}(\vep)=4$
   and $\varepsilon =0$ denotes the chemical potential. 
Figure \ref{dos_site_fig} depicts $D_{\alpha}(\vep)$ 
   close to the Fermi energy. 
The local density of states, $D_{\alpha}(\vep)$, 
   for small $\varepsilon$  is the largest at C 
   site corresponding to the electron rich site, and 
     is the smallest  at B site corresponding to 
     the hole (or charge) rich site \cite{komaba}. 
We notice that the above result is consistent with that of the 
   experiment \cite{kakiuchi}.

The relative magnitude of $D_{\alpha}(\vep)$ 
 is understood as follows. 
From  \S \ref{tilted_weyl_equation}, 
  we obtain that $d_{\alpha\gamma}(\mib{k})$ depends on only 
  $\theta(=\tan^{-1}[k_y/k_x])$ 
    in the vicinity of the Dirac point, 
  and that the energy band is described by eq. \refeq{eq:16}. 
 Then 
  $D_{\alpha}(\vep)$ $(\vep > 0)$ can be rewritten as 
\begin{align}
D_{\alpha}(\vep)
&=
\frac{|\varepsilon|}{2\pi^2}\int^{2\pi}_{0}{\rm d} \theta
\int_0^\infty  {\rm d}r \; r
|d_{\alpha1}(\theta)|^2
 \delta (\varepsilon - r/F(\theta))
\nonumber \\
&=\frac{|\varepsilon|}{2\pi^2}\int^{2\pi}_{0}{\rm d} \theta
|d_{\alpha1}(\theta)|^2(F(\theta))^2, 
\label{dos_fe}
\end{align}
  and 
\begin{align}
\sum^{4}_{\alpha=1}D(\vep)
&=\frac{|\varepsilon|}{2\pi^2}\int^{2\pi}_{0}{\rm d} \theta
(F(\theta))^2, 
\label{dos_fe_all}
\end{align}
  in the low energy region. 
The integration of eq. \refeq{dos_fe} is dominant for $\theta\sim0$, 
   since $F(0)$ is larger than $F(\pi)$ as shown in 
    Figure \ref{ellipse_fig}. 
Further, from Figure \ref{zv_theta} (a), 
  we see that the component of the eigenvector satisfies a condition 
  $|d_{41}(0)|>|d_{11}(0)|=|d_{21}(0)|>|d_{31}(0)|$. 
These facts give a conclusion 
  that the contribution of the C site (electron rich site) 
   to the density of state 
   is larger than that of the B site (hole rich site). 
It is also found, within the present scheme of eq.~(\ref{h_eff2}),  
  that  the density of states of $\vep < 0$  is the same as that of 
   $\vep > 0$, from the symmetry of the energy band, 
   $\Delta E_1(\mib{k})=-\Delta E_2(-\mib{k})$ 
   (see eq. \refeq{linear_dispersion}). 
The underlying symmetry is due to the transformation 
  property of the effective Hamiltonian under space 
  inversion \cite{montambaux}. 
We note that eq. \refeq{dos_fe_all} is the same as 
   $|\vep|/(2\pi^2v^{*2}_F)$ obtained in Ref. \cite{montambaux}. 
 The qunatity  $v^{*}_F$ is the renormalized velocity, and is estimated 
  as 
  $v^{*}_F=0.0097$ for the choice of the present parameters
   where  the corresponding 
    four effective velocities in Ref. \cite{montambaux} 
    are given  as 
   $w_{0x}=-0.039$, $w_{0y}=0.008$, 
   $w_{x}=0.052$, and $w_{y}=0.044$. 
For small $|\vep|$, the total density of states shown in 
  Figure \ref{dos_site_fig} coincides well with both 
   eq. \refeq{dos_fe_all} and $|\vep|/(2\pi^2v^{*2}_F)$.

Finally, the magnetic properties are examined. 
The susceptibility is calculated  on 
   the basis of four sites in a unit cell. 
The susceptibility corresponding to  the response  between 
 the $\alpha$ site and the $\beta$ site  is described as,  
\begin{align}
&\chi^0_{\alpha\beta}(\mib{q},\omega)=\nonumber\\
&-\frac{1}{N}\sum^{}_{\mib{k}\gamma\gamma'}
d_{\alpha\gamma}(\mib{k}+\mib{q})
d^*_{\beta\gamma}(\mib{k}+\mib{q})
d_{\beta\gamma'}(\mib{k})
d^*_{\alpha\gamma'}(\mib{k})\nonumber\\
&\times\frac{f(\xi_{\gamma}(\mib{k}+\mib{q}))-f(\xi_{\gamma'}(\mib{k}))}
{\xi_{\gamma}(\mib{k}+\mib{q})-\xi_{\gamma'}(\mib{k})-\omega-\rmi\delta},
\label{eq:irreducible}
\end{align}
   where $\rmi\delta$ ($\delta > 0$)
  is a infinitesimally small imaginary part. 
Using equation (\ref{eq:irreducible}),
 the magnetic responses of both the local susceptibility and 
  the local NMR relaxation rate \cite{moriya,suzumura_nmr} 
   at the $\alpha$ site are obtained as
\begin{align}
\chi_{\alpha}&=\sum^{}_{\beta}
  \mathrm{Re}(\chi^0_{\alpha\beta}(\mib{0},0))=
    \int^{\infty}_{-\infty}d\vep
     D_{\alpha}(\vep)\left(-\frac{\partial f}{\partial\vep}\right),
\label{chi_static}\\
\left(\frac{1}{T_1}\right)_{\alpha}&=T\lim_{\omega\rightarrow0}
  \sum^{}_{\mib{q}}
   \frac{\mathrm{Im}(\chi^0_{\alpha\alpha}(\mib{q},\omega))}{\omega}
\nonumber\\
&=\pi T
  \int^{\infty}_{-\infty}d\vep
   \left(D_{\alpha}(\vep)\right)^2
    \left(-\frac{\partial f}{\partial\vep}\right),
\label{t1t_eq}
\end{align}
  respectively.

\begin{figure}
\begin{center}
\resizebox{0.35\textwidth}{!}{
\includegraphics{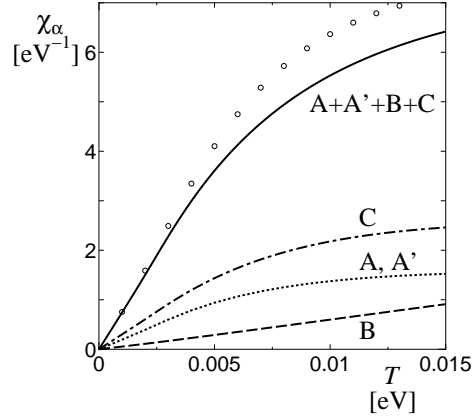}
}
\caption{
Temperature dependence of $\chi_{\alpha}$ ($\alpha=A, A', B, C$). 
  Dotted line, dashed line, and dot-dashed line correspond to 
    A (A'), B, and C sites, respectively. 
The total value of $\chi_{\alpha}$ is plotted by the solid line
 where the open circle is calculated using eq.~(\ref{te_temp}). 
}
\label{chia}
\end{center}
\end{figure}

\begin{figure}
\begin{center}
\resizebox{0.35\textwidth}{!}{
\includegraphics{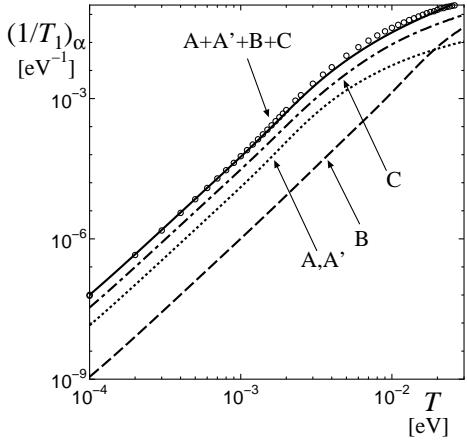}
}
\caption{
Temperature dependence of $(1/T_1)_{\alpha}$. 
Dotted line, dashed line, and dot-dashed line represent 
  $(1/T_1)_{\alpha}$ for A (A'), B, and C sites, respectively. 
The total value of $\chi_{\alpha}$ is plotted by the solid line
 where the open circle is calculated by using eq.~(\ref{te_temp}).  
 }
\label{t1t}
\end{center}
\end{figure}

The local magnetic susceptibility and the local NMR relaxation rate 
   for each site are shown as a function of temperature 
   in Figures \ref{chia} and \ref{t1t}, respectively. 
The susceptibility shows a relation 
  $\chi_C>\chi_A=\chi_{A'}>\chi_B$, which agrees qualitatively 
  with that of the experiment \cite{takahashi_pc}. 
The open circles denotes the total one, which is calculated using 
 the temperature dependence of the chemical potential 
 and $\lrangle{n_{\alpha}}$ 
 (eqs. \refeq{eigen_equation}-\refeq{nalpha}) and 
  the transfer energy given by  
\begin{align}
t_{L}(T)&=\frac{t_L(300\mathrm{K})-t_L(8\mathrm{K})}{292\mathrm{K}}
(T-300\mathrm{K})+
t_L(300\mathrm{K}). 
\label{te_temp}
\end{align}
  The data of $t_L(300\mathrm{K})$ and $t_L(8\mathrm{K})$
  ($L=a1,\cdots,a3$, $b1,\cdots,b4$, $a1',a3',a4'$) 
  are given in Ref. \cite{kino_1}. 
The deviation is visible for $T > 0.005$ eV. 
In Figures \ref{chia} and \ref{t1t}, 
   reflecting the density of states,  
   $\chi_{4}$ and $(1/T_1)_{4}$ (C site) are the largest among
   those of the four sites. 
It is found that, at low temperature, 
  $\chi_{\alpha}$ and $(1/T_1)_{\alpha}$ 
  are proportional to $T$ and $T^3$, respectively, 
   due to $D_{\alpha}(\vep)\propto|\vep|$. 
Actually,  
  eqs.~\refeq{chi_static} and \refeq{t1t_eq} at low temperature
   are calculated as 
  $\chi_{\alpha} = 2 ({\rm ln}2) K_{\alpha}T$ 
   and 
   $(1/T_1)_{\alpha} =  (\pi^3/3)K_{\alpha}^2 T^3$
   where $D(\varepsilon) = K_{\alpha} |\varepsilon|$ from 
   eq. (\ref{dos_fe}), 
   where 
   $K_A \simeq  137.1$, $K_B \simeq 38.9$ and $K_C \simeq 212.9$.

\section{Summary and discussion}
\label{summary}
We examined the ZGS close to the Dirac point, 
   and local magnetic   properties at low temperature.  
Each electron in the unit cell has respective role to form  a Dirac 
  particle which is 
   different from that of the graphene. 
The C  and  B sites  give the largest and smallest density of  
  states respectively, and are peculiar for the present salt 
   $\alpha$-(BEDT-TTF)$_2$I$_3$ while the A and A' sites have a common 
   feature with the graphene. The temperature dependence of magnetic 
    susceptibility and NMR relaxation rate exhibit the  power law at low 
    temperature, which can be interpreted in terms of the density of states
     close to the Fermi surface.

Here  we note that  the detail of the ZGS  depend  on 
    the choice of parameters as transfer energy, interaction and pressure.
 There are two kinds of conditions for the existence of the ZGS.
First one is  the  contact point, which  
is satisfied for most of $\alpha$-type BEDT-TTF salt, 
  and may arise from a property of 
  an inversion symmetry of the four sites in the unit cell. 
The second one is that 
 the contact point exists on the Fermi energy (i.e, the disappearance of 
     hole or electron pockets).  
The ZGS has been obtained in the following cases. 
(i) Using the data of the X-ray experiment at room 
    temperature \cite{kondo} and $U =0.4$, $V_a=0.17$, $V_a=0.05$, 
    the ZGS occurs for $P_a>4.3$ kbar, whereas 
    the charge ordered insulating (metallic) state is obtained for 
    $0$ kbar $<P_a<3.3$ kbar ($3.3$ kbar$<P_a < 4.3$ kbar) 
    \cite{kobayashi_2004_sc}. 
(ii) In the absence of interaction, the ZGS (metallic) state 
    is obtained for  $P_a > 3$ kbar (3 kbar$< P_a$) \cite{katayama_zgs}. 
(iii) 
  For the small magnitude of interaction,
  the ZGS is obtained 
      due to reducing the pockets while  the large interaction  
       destroys the ZGS  due to the formation of the charge ordered (CO) state. 
(iv) Further, 
         the ZGS appears under higher 
     hydrostatic pressure ($12.5$ kbar  $< P$) \cite{katayama2008_sdw}.

We note the following relation  between 
   the charge disproportionation and the magnetic susceptibility
    in the ZGS. 
Using  parameters  at 
   $P_a=$10 kbar and room temperature \cite{kondo}, 
  the  charge disproportionation is given by 
  $\lrangle{n_C} > \lrangle{n_A} =\lrangle{n_{A'}} > \lrangle{n_B}$ 
    while  the susceptibility ($ \propto T$)  
       is obtained  as $\chi_A (=\chi_{A'}) >  \chi_C >  \chi_B$. 
This is contrast to the case for transfer energies calculated 
  from the first principle calculation at 8K and 
    at ambient pressure \cite{kino_1}, where 
         $\lrangle{n_C} > \lrangle{n_A} =\lrangle{n_{A'}} > \lrangle{n_B}$ 
     and  $\chi_C > \chi_A  =  \chi_{A'} >  \chi_B$.  
For the present choice of the transfer energy \cite{kino_1},   
    the ZGS (CO) state  is obtained for $x <1.03$ $(x>1.03)$
       with  $U = 0.4 x$, $V_a=0.17 x$ and $V_b=0.05 x$. 
Thus, we propose the following variation of interaction in order to 
  explain the experiment that 
  the CO state is obtained at ambient pressure and the ZGS is obtained for 
  $P_a>4.3$ kbar. 
The parameter of interaction at $P_a=0$ kbar is taken as $x>1.03$ and 
  that of $P_a>4.3$ kbar is expected as $x<1.03$, 
since the effect of pressure increases the band width and 
  decrease the relative magnitude of interactions. 

\begin{figure}
\begin{center}
\resizebox{0.35\textwidth}{!}{
\includegraphics{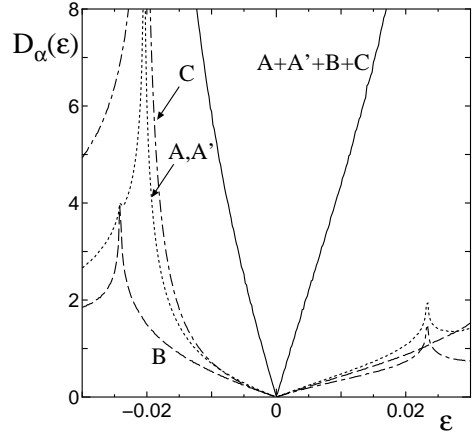}
}
\caption{
Density of states at the ambient pressure and at $T=8$ K where 
  interactions are the same as Figure \ref{dos_site_fig}. 
}
\label{dos_kondo}
\end{center}
\end{figure} 
Finally we discuss the state, which is derived from 
  the transfer energies obtained from the experimental data 
    at $T=8$ K and at ambient pressure \cite{kondo}. 
In this case, the ZGS emerges even at ambient pressure. 
Figure \ref{dos_kondo} shows the density of states 
  under the ambient pressure where the parameters of interactions are 
    the same as those of Figure~2.  
The charge disproportionation is obtained as
  $\lrangle{n_C}>\lrangle{n_A}=\lrangle{n_{A'}}>\lrangle{n_B}$ 
   which is qualitatively the same  
    as the first principle calculation.\cite{kino_1}
However, compared with Figure~\ref{dos_site_fig}, 
   the difference of each of $D_{\alpha}(\vep)$ is very small 
    in the vicinity of the Fermi energy. 
This implies that  the similarity between  $\chi_{\alpha}$ 
  (corresponding to  $D_{\alpha}(\vep)$ ) and $\lrangle{n_{\alpha}}$ 
  is not necessary for the existence of the ZGS. 
We also note that the node found in Figure~\ref{zv_theta} 
  also exists in the case of 
  Figure~\ref{dos_kondo} although the location of the node is different. 
Thus the role of respective sites seems 
  to be a common feature for the ZGS of the system 
     with four sites in the unit cell. 
Further we comment on  a large Van Hove singularities, which exist for 
  $\varepsilon > 0$  in Figure~\ref{dos_site_fig},
    and for  $\varepsilon < 0$ in  Figure~\ref{dos_kondo}.  
These singularities give opposite results for temperature effect, e.g. 
    with increasing temperature, 
    the chemical potential decreases for  Figure~\ref{dos_site_fig}  but 
   increases  for Figure~\ref{dos_kondo}
         with increasing temperature.
Thus the case of Figure \ref{dos_kondo} needs further consideration 
   in order to explain a fact\cite{kobayashi_hall}  that  
    the behavior of the Hall coefficient suggests 
     the monotonic decrease of chemical potential
       with increasing temperature.

\section*{}
We are grateful to Dr. Y. Takano and Prof. T. Takahashi for  
  encouraging the motivation of our work 
   and also for useful discussions. 
  S. K. acknowledges the financial support of Research Fellowship for 
  Young Scientists from Japan Society for the Promotion of Science (JSPS). 
This work was also financially supported by a 
   Grant-in-Aid for Scientific Research on 
  Priority Areas of Molecular Conductors (No. 15073103) from 
    the Ministry of Education, 
     Culture, Sports, Science and Technology, Japan. 

\appendix
\section*{Appendix: A 
constant energy curve for the general tilted Weyl equation}
In this appendix, we examine the relation between 
  the linear dispersion calculated from the tilted Weyl equation and 
  the ellipse as shown in Figure \ref{ellipse_fig}. 
When we put $\lambda_{+}(\mib{k})=\omega$ in 
  eq. \refeq{linear_dispersion}, we obtain the 
  general equation of the ellipse centered on $(X_0,Y_0)$ 
  as follows: 
\begin{align}
&\frac{V_x^2-v_{0x}^2}{R^2}( k_x-X_0)^2+
\frac{V_y^2-v_{0y}^2}{R^2}( k_y-Y_0)^2\nonumber\\
&+
\frac{V_{xy}^2-v_{0x}v_{0y}}{R^2}
(k_x-X_0)(k_y-Y_0)=1,
\end{align}
where
\begin{align}
R=\omega\left[
\frac{V_{x}^2V_{y}^2-V_{xy}^4}
{(V_{x}^2-v_{0x}^2)(V_{y}^2-v_{0y}^2)-
(V_{xy}^2-v_{0x}v_{0y})^2}
\right]^{1/2}.
\end{align}
The coordinates of the center $(X_0,Y_0)$ are given in the form: 
\begin{align}
\frac{X_0}{\omega}&=\frac{v_{0y}V_{xy}^2-v_{0x}V_{y}^2}
{(V_x^2-v_{0x}^2)(V_y^2-v_{0y}^2)-(V_{xy}^2-v_{0x}v_{0y})^2}\; ,
\nonumber\\
\frac{Y_0}{\omega}&=\frac{v_{0x}V_{xy}^2-v_{0y}V_{x}^2}
{(V_x^2-v_{0x}^2)(V_y^2-v_{0y}^2)-(V_{xy}^2-v_{0x}v_{0y})^2}\; .
\end{align}
Carrying out the coordinate rotation,
\begin{align}
&\left(
\begin{array}{c}
k'_{x}\\k'_{y}
\end{array}
\right)=
\left(
\begin{array}{cc}
\cos\phi&\sin\phi\\
-\sin\phi&\cos\phi
\end{array}
\right)
\left(
\begin{array}{c}
 k_x-X_0\\
 k_y-Y_0
\end{array}
\right)\nonumber\\
&\phi=\left\{
\begin{array}{l}
\dfrac{1}{2}
\tan^{-1}\left(
\dfrac{2(V_{xy}^2-v_{0x}v_{0y})}
{(V_x^2-v_{0x}^2)-(V_y^2-v_{0y}^2)}
\right)
\\
\qquad\qquad\qquad\qquad\qquad
(V_x^2-v_{0x}^2<V_y^2-v_{0y}^2)
\\
- \dfrac{\pi}{2}
+\dfrac{1}{2}
\tan^{-1}\left(
\dfrac{2(V_{xy}^2-v_{0x}v_{0y})}
{(V_x^2-v_{0x}^2)-(V_y^2-v_{0y}^2)}
\right)\\
\qquad\qquad\qquad\qquad\qquad
(V_x^2-v_{0x}^2>V_y^2-v_{0y}^2)
\end{array}
\right.,
\end{align}
  we obtain the standard form of the ellipse (Fig. \ref{ellipse_fig}) 
  given by 
\begin{align}
\frac{k_x^{2}}{a^2}+\frac{k_y^{2}}{b^2}=1,
\end{align}
where the long and short axes are given as 
\begin{align}
a&=\frac{\sqrt{2}R}
{\left[V_x^2-v_{0x}^2+V_y^2-v_{0y}^2-
\sqrt{D}\right]^{1/2}}\; ,
\nonumber\\
b&=\frac{\sqrt{2}R}
{\left[V_x^2-v_{0x}^2+V_y^2-v_{0y}^2+
\sqrt{D}
\right]^{1/2}}\nonumber\\
D&=
[(V_x^2-V_y^2)-(v_{0x}^2-v_{0y}^2)]^2+
4[V_{xy}^2-v_{0x}v_{0y}]^2.
\end{align}



\end{document}